\documentclass[12pt]{revtex4}

\usepackage{amsfonts}
\usepackage{latexsym}
\usepackage{amssymb}
\usepackage{epsfig}
\begin{document}
\oddsidemargin=0.8cm
\newtheorem{axiom}{Theorem}[section]
\newtheorem{definition}{Definition}[section]
\newtheorem{proposition}{Proposition}[section]
\newtheorem{lemma}{Lemma}[section]
\def\bib{\bibitem}
\def\be{\begin{equation}}
\def\ee{\end{equation}}
\def\ba{\begin{eqnarray}}
\def\eea{\end{eqnarray}}
\def\df{\stackrel{\rm def}{=} }
\renewcommand{\theequation}{\arabic{section}.\arabic{equation}}
\def\R{\mathbb{R}}
\def\d{{\cal D}}

\title{Existence of the Bogoliubov $S(g)$ Operator for the 
$(:\phi^4:)_2$ Quantum Field Theory}
\author{\bf Walter F. Wreszinski, Luiz A. Manzoni and Oscar Bolina$^a$}
\affiliation{Instituto de F\'{\i}sica\\
Universidade de S\~ao Paulo\\
Caixa Postal 66318 \\
05315-970 -- S\~ao Paulo, SP\\
Brasil/Brazil}
\date{\today}
\maketitle
\vskip 1 cm
\noindent
\begin{center}
{\bf Abstract}
\end{center}
\noindent
We prove the existence of the Bogoliubov $S(g)$ operator for the 
$(:\phi^4:)_2$ quantum field theory for coupling functions $g$ of compact support in space and time. The construction is 
nonperturbative  and relies on a theorem of Kisy\'nski. It implies 
almost automatically the properties of unitarity and causality for 
disjoint supports in the time variable.
\vskip .5 cm
\noindent
\vfill
\hrule width2truein \smallskip {\baselineskip=10pt \noindent
{\small {\bf E-mails:} wreszins@fma.if.usp.br, lmanzoni@fma.if.usp.br, 
oscar@math.ucdavis.edu \\
$^a$ Present address: Department of Mathematics, University of California, Davis, CA 95616-8633, USA}
\par
}
\section{Introduction and Summary}
\label{intro}

Recent progress in perturbative quantum field theory
for the St\"uckelberg-Bogoliu{\-}bov-Epstein-Glaser $S(g)$ operator 
\cite{BLT75, EGl73} in nonabelian gauge theories \cite{Sch01} (see also 
\cite{DFr99}), revived interest in a long-standing problem: is it
possible to construct $S(g)$ {\it nonperturbatively} in
quantum field theory? This question is of obvious relevance to theories 
where the (dimensionless) coupling constant is large ($\gtrsim 1$) -- e.g. 
strong interactions -- for which perturbation theory is not expected to be 
asymptotic.

For certain super-renormalizable theories -- the $(:P(\phi ):)_2$ theories 
-- there exists, for weak coupling, a construction of the true 
(LSZ-Haag-Ruelle) 
scattering operator, due to  Osterwalder and S\'eneor \cite{OSe76} and 
Eckmann, Epstein and Fr\"ohlich \cite{EEF76}, one of the crowning 
achievements of constructive quantum field theory which started with the pioneering work on the particle structure of weakly coupled $P(\phi )_2$ model by J. Glimm, A. Jaffe and T. Spencer \cite{GJS74} . The method of proof was, 
however, perturbative: the perturbation series for the scattering operator 
was shown to be asymptotic.

In contrast to the true scattering operator, $S(g)$ is, in perturbation 
theory, the generating functional for the time-ordered products of Wick 
polinomials. However, on the basis of \cite{EGl76} one might expect that, 
in the present massive case, defining 
$$g_{\varepsilon}(x)\equiv g(\varepsilon x)\; 
;\hspace{1.0cm}g\in {\cal S} (\mathbb{R}^2)$$

\noindent
the (adiabatic) limit
\be
S\Psi \equiv \lim_{\varepsilon \rightarrow 0}\; S(g_{\varepsilon})\Psi
\label{adiab}
\ee
\noindent
exists, $\forall \;\Psi \in {\cal D}$, where ${\cal D}$ is a 
Poincar\'e-invariant dense set in Fock space ${\cal F}$. Thus we expect 
that the physical $S$-matrix elements are obtainable as 
\be
(\Phi, S\Psi)\equiv \lim_{\varepsilon \rightarrow 0}\; \; (\Phi , 
S(g_{\varepsilon})\Psi)\; , \label{adiab2}
\ee
\noindent
with $\Phi \in {\cal F}$, $\Psi \in {\cal D}$, where $g(0)>0$ should be 
identified  with the coupling constant. In \cite{DFr99} an algebraic 
construction of the adiabatic limit was performed for 
perturbative QED. 

A natural nonperturbative approach to construct 
$S(g)$ for the $(:\phi^4:)_2$ theory (and hopefully for any 
super-renormalizable QFT) consists in proving the existence of a 
(unique) solution of the evolution (propagator) equation ($\hslash =1$)
\be
i\frac{\partial U(t,s)}{\partial t}\Psi = \tilde{H}(t)U(t,s)\Psi\; ,
\label{eqprop}
\ee
with
\be
\tilde{H}(t) \equiv H_g(t) + M {\bf 1}\; ,
\ee
where $M$ is a constant introduced in order to make $\tilde{H}(t)$ a positive operator (see section \ref{existence}) and 
\be
H_g(t) \equiv H_0 + V_g(t)\; .
\label{hg}
\ee
In (\ref{eqprop}) $U(t,s)$ is a two-parameter family of unitary operators on 
(symmetric) Fock space ${\cal F}$. $H_0$ is the free field Hamiltonian 
corresponding to a zero-time scalar field $\phi (x,0)$ of mass $m$ 
\cite{GJa70, GJa72}, and, formally, for 
\be
g\in {\cal D}(\mathbb{R}^2)\; ; \hspace{1.0cm} g\geq 0 \; ,
\label{g}
\ee
\noindent
let
\be
V_g(t)=\int\; dx\; g(x,t):\phi^4(x,0):\; .
\label{vg}
\ee
Above, ${\cal D}$ denotes the Schwartz space of infinitely differentiable 
functions of compact support. The operators in (\ref{eqprop}) are expected 
to satisfy the propagator conditions:
\ba
&&U(t,s)U(s,r)=U(t,r)\; ,\hspace{1.0cm}-\infty <r\leq s\leq t<\infty 
\; ,\nonumber \\ \label{propcon}\\
&&U(t,t)= {\bf 1}\; ,\hspace{3.3cm} \forall \;t \in \mathbb{R}\; . \nonumber 
\eea
\noindent
The vector $\Psi$ is supposed to belong to the domain $D (\tilde{H}(s))$ (dense 
in ${\cal F}$) such that 
\be
U(t,s)D( \tilde{H}(s) ) \subset D( \tilde{H}(t) )\; .
\label{udcon}
\ee
Above and elsewhere in this paper $D(A)$ denotes the domain of 
an operator $A$. 

Under assumptions (\ref{eqprop}) and (\ref{udcon}), defining the ``Dirac 
(or interaction) picture propagator'' by
\be
U^D(t,s)\equiv e^{i(H_0+M)t}U(t,s)e^{-i(H_0+M)s}\; ,
\label{propdir}
\ee
\noindent
it follows that
\be
i \frac{ \partial U^D(t,s)}{\partial t}\Psi = H_g^D(t) U^D (t,s)\Psi \; ,
\label{eqpropd}
\ee
\noindent
for $\Psi \in e^{i(H_0+M)s}D(H_g(s))$, which is a dense set in ${\cal F}$ 
for every $s$, where
\be
H_g^D(t)\equiv e^{iH_0t}V_g(t)e^{-iH_0t}\; .
\label{hd}
\ee
\noindent
One may then define
\be
S(g)\equiv s-\lim_{\scriptscriptstyle {{ t\rightarrow + \infty} \atop 
{s\rightarrow - \infty } } } U^D(t,s) \; ,
\label{sg}
\ee
\noindent
if the above limit exists; $S(g)$ is expected to
satisfy
\begin{itemize}

\item[$(i)$] $S(g)^{-1}=S(g)^*\;$  ({\it unitarity});

\item[$(ii)$] $S(g_1 +g_2)= S(g_1)S(g_2)$ if

     \begin{itemize}
     
     \item[$(ii.a)$] supp $g_1 >$ supp $g_2$

     \item[] and/or

     \item[$(ii.b)$] supp $g_1 \sim$ supp $g_2$  ({\it causality})

     \end{itemize} 

\noindent
where ``$\sim$'' means ``spacelike to'', i.e., $(x-y)^2= 
(t_1-t_2)^2-(x_1-x_2)^2<0$, $\forall \; (t_1,x_1) \in {\rm supp}\; g_1 $ 
and $\forall \; (t_2,x_2) \in {\rm supp}\; g_2 $;

\item[$(iii)$] There exists a unitary representation $U(a, \Lambda )$ of 
the Poincar\'e group on ${\cal F}$ -- the scalar field representation of 
mass $m$ -- such that
$$U(a, \Lambda ) S(g)U(a, \Lambda )^{-1} = S(\{ a, \Lambda \} g)\; ,$$ 
\noindent 
where
$$(\{ a, \Lambda \} g)(x) =g(\Lambda^{-1}(x-a))\; $$ 
({\it Lorentz covariance}).

\end{itemize} 

The main difficulty to proving (\ref{eqprop})-(\ref{udcon}) is  
that $D(H_g(t))$ is, for each $g \in {\cal D}(\mathbb{R}^2)$, 
time-dependent. In section 
\ref{existence} we state the basic existence theorem we employ, 
which is due to Kisy\'nski \cite{Kis64} (see also \cite{Sim71}). In 
section \ref{prova} we prove our central existence theorem for $S(g)$, as 
well as properties $(i)$ and $(ii.a)$. In section \ref{kis} we provide a 
brief summary of the remarkable results of \cite{Kis64}, establishing 
a concrete link between them and our conditions in section \ref{prova}. We 
leave the conclusion and open problems to section \ref{conclusion}. Appendix A summarizes some of the basics elements of the construction of \cite{Kis64} and \cite{Yos80} for the convenience of the reader.

\section{The Basic Existence Theorem}
\label{existence}
\setcounter{equation}{0}

The Hamiltonian of the $(:\phi^4:)_2$ theory \cite{GJa68} is given by (\ref{hg}), where
\be
H_0=\int \omega (k)a^*(k)a(k) dk\; ,
\label{H0}
\ee
with
\be
\omega (k)=\left( k^2+m^2\right)^{\frac{1}{2}}\; ,
\label{omega}
\ee
\noindent
is the free field Hamiltonian on symmetric Fock space ${\cal F}$, with 
\be
[a(k),a^*(k')]=\delta (k-k')\; .
\label{com}
\ee
The self-interaction $V_g$ is given by (\ref{vg}), with the $t=0$ scalar 
free field of mass $m$:
\be
\phi (x)=\frac{1}{(4\pi )^{\frac{1}{2}}}\int e^{-ikx}\left[ a^*(k)+a(-k)\right]\omega (k)^{-\frac{1}{2}}dk  \; .
\label{phit0}
\ee
Thus $V_g$ may be written \cite{GJa68}
\ba
V_g(t)&=& \sum_{j=0}^4 {4 \choose j} \int a^*(k_1)\cdots
 a^*(k_j)a(-k_{j+1})\cdots a(-k_4) \nonumber  \\  \\
&\times&\tilde{g}({\scriptstyle \sum\limits_{i=1}^4
 } k_i , t)\prod_{i=1}^4\omega (k_i)^{-\frac{1}{2}} dk_i\;\nonumber 
 , \nonumber
\label{HI}
\eea
where 
\be
\tilde{g}(k,t) \equiv \int dx \; e^{ikx} g(x,t) \; .
\label{tg}
\ee

The number operator $N$ is defined by 
\be
N=\int dk \;a^*(k)a(k) \; .
\label{N}
\ee

By \cite{GJa68} (Lemma 2.2)
\be
\left\| (N + {\bf 1})^{-\frac{j}{2}} \; V_g(t) \; 
(N + {\bf 1})^{-\frac{4 - j}{2}} \right\| \leq \; {\rm const.}
\; \left\| W \right\|_{L^2} \; , \hspace{1.0cm} |j| \leq 4
\label{gjdes}
\ee
where
\be\label{w}
W(k,t)\equiv \tilde{g}({\scriptstyle \sum\limits_{i=1}^4} k_i , 
t)\prod_{i=1}^4\omega (k_i)^{-\frac{1}{2}} \; .
\ee
The above mentioned lemma just uses the Fock space definitions of the 
creation and annihilation operators and the Schwartz inequality. We 
need two theorems due to Glimm and Jaffe, which we state as adapted to 
our case:
\begin{axiom}\label{sa}{\rm \cite{GJa68}} {\bf (a)} $H(t)$ is 
self-adjoint on the domain 
\be
D(H(t))=D(H_0) \cap  D(V_g(t))\; ,\label{DHtotal}
\ee
\noindent
where $D(V_g(t))$ is the domain of the unique self-adjoint closure of 
$V_{g}(t)$ on the domain
\be\label{dom}
D_{0}=\bigcap_{n=0}^{\infty} D(H^{n}_{0}).
\ee
{\bf (b)}  $H(t)$ is essentially self-adjoint on $D_{0}$.
\end{axiom}
\begin{axiom}\label{posit} {\rm \cite{GJa72}}
For each $g \in {\cal D}(\mathbb{R}^{2})$, there exists $0 < M_{g} < 
\infty$ such that
\be\label{hinf}
H_g(t) \geq -M_{g} {\bf 1}
\ee
as a bilinear form on $D_{0} \times D_{0}$.
\end{axiom}

By theorem \ref{posit} and {\bf (b)} of theorem \ref{sa}, $H(t)$ is a 
semi-bounded self-adjoint operator, and thus defining
\be\label{mmc}
M=M_{g}+c,
\ee
for some $c > 0$, then
\be\label{hmc}
\tilde{H}(t)=H_g(t)+M{\bf 1} \geq c {\bf 1} 
\ee
is a positive self-adjoint operator. Let ${\cal F}_{+2}=D(H_{0})$
endowed with the Hilbert space structure given by
\be\label{f+2}
f_{+2}(x,y)=\langle \left (H_{0}+1 \right )x, \left (H_{0}+1 \right )y 
\rangle
\ee
and denote $\sqrt{f_{+2}(x,x)}$ by $||x ||_{+2}$. By the Riesz lemma
we may associate ${\cal F}_{+2}$ and the space ${\cal F}_{-2}$ of 
continuous conjugate linear functions on ${\cal F}_{+2}$. While we
consider ${\cal F}$ isomorphic to its conjugate dual space ${\cal F}^{*}$, 
the isomorphism being the identity, the isomorphism of ${\cal F}_{+2}$
with ${\cal F}_{-2}$ is given by the operator $\left (H_{0}+1 \right 
)^{2}$, because
\[
||v||_{-2}=\sup{ \left \{ |\langle w, v \rangle | ~:~ ||w ||_{+2} \leq 1 
\right \} }.
\]
Since $f_{+2}(x,y)=\langle x, \left (H_{0}+1 \right )^{2}y \rangle$, we
have 
\begin{eqnarray}
\left\| \left (H_{0}+1 \right )^{2}y \right\|_{-2}&=&
\sup{ \left \{ | \langle w, \left (H_{0}+1 \right )^{2}y \rangle | ~:~ 
||w ||_{+2}=\sqrt{\langle w, \left (H_{0}+1 \right )^{2}w \rangle}  \leq 1 
\right \} } \nonumber\\
&=&\left\| \left (H_{0}+1 \right )y \right\| =||y||_{+2}, \nonumber
\end{eqnarray}
from which we also have, for $y \in {\cal F}$,
\be\label{n-2}
||y||_{-2}=||\left (H_{0}+1 \right )^{-1}y ||,
\ee
which explains the notation ${\cal F}_{-2}$. Clearly $||x || \leq || x ||_{+2}$ for $x \in {\cal F}_{+2}$, and by 
(\ref{n-2}), $||y ||_{-2} \geq || y ||$ for $y \in {\cal F}$. Thus, under 
the above conditions:
\be\label{escala}
{\cal F}_{+2} \subset {\cal F} \subset {\cal F}_{-2}.
\ee

A bounded operator {\bf B} from ${\cal F}_{+2}$ to ${\cal F}_{-2}$
is thus such that, for some constant $c$,
\be\label{B}
||{\bf B} \psi ||_{-2} \leq c || \psi ||_{+2}\;\;\;\;\;\;\;\; \psi 
\in {\cal F}_{+2},
\ee
or, by (\ref{f+2}) and (\ref{n-2}),
\be\label{HB}
||\left (H_{0}+1 \right )^{-1}{\bf B} \psi|| \leq c || \left (H_{0}+1 
\right )\psi || \;\;\;\;\;\;\;\; \psi \in {\cal F}_{+2},
\ee
or
\be\label{HBH}
||\left (H_{0}+1 \right )^{-1}{\bf B}\left (H_{0}+1 \right )^{-1} 
\phi|| \leq c || \phi || \;\;\;\;\;\;\;\; \phi \in 
{\cal F}.
\ee
Now, by (\ref{hmc}), we may define ${\tilde{H}}(t)^{1/2}$, and, by 
(\ref{gjdes}) for $x \in {\cal F}_{+2}$, the closed sesquilinear form
\be
\label{forma}
S(x,y)= \langle \tilde{H}(t)^{1/2}x, \tilde{H}(t)^{1/2}y \rangle
\ee
which is, by the form representation theorem \cite{Far75}, 
the form of the operator ${\tilde{H}}(t)$. In section \ref{prova} we show 
the explicit connection of (\ref{forma}) to the basic theorem of 
Kisy\'nski \cite{Kis64}, which we state in the form of theorems II.23 and 
II.24 of \cite{Sim71}, with slight changes.

In the theorem stated below, ${\cal F}_{\pm 2}$ have been defined in (\ref{f+2}) - (\ref{escala}).

\begin{axiom}\label{texist}
Let (\ref{escala}) hold and $\tilde{H}(t)$ $(-T \leq t \leq S)$ be
a one-parameter family of strictly positive (i.e. satisfying (\ref{hmc}))
self-adjoint operators on ${\cal F}$. Suppose that 
$\tilde{H}(t): {\cal F}_{+2} \rightarrow {\cal F}_{-2}$ are
bounded and twice differentiable, with a continuous second derivative, in 
the $|| \cdot ||_{-2,2}-$norm (\ref{B}). Then there exists a 
two-parameter family $U(t,s)$ of unitary propagators satisfying 
(\ref{eqprop}), (\ref{propcon}) and (\ref{udcon}). 
\end{axiom}


\section{The central existence theorem}\label{prova}
\setcounter{equation}{0}

We now use theorem \ref{texist} in order to prove our main 
\begin{axiom}\label{main}
The $\left (: \phi^{4} : \right )_{2}$ theory, as defined by (\ref{hg}), 
(\ref{g}), (\ref{vg}), (\ref{H0}) and (\ref{omega}), satisfies a stronger 
condition than the hypothesis of theorem \ref{texist}: $H_{g}( \cdot )$ 
is infinitely differentiable as an operator from ${\cal F}_{+2}$ to 
${\cal F}_{-2}$. 
\end{axiom}
In order to prove theorem \ref{main} we first show a useful auxiliary result.
\begin{lemma}\label{wg}
Let $W$ be defined by (\ref{w}). Then there exists $r > 1$ such that
\be\label{w2}
|| W( \cdot , t) ||_{2} \leq  const. \; || g(\cdot , t) ||_{r}
\ee
where 
\be\label{gr}
|| g(\cdot , t) ||_{r}=\left (\int_{-\infty}^{+\infty} 
dk | \tilde{g}(k,t)|^{r} \right )^{1/r}.
\ee
\end{lemma}
{\bf Proof.} We have
\ba
|| W(\cdot , t) ||^{2}_{2}&=&\int_{-\infty}^{+\infty} dk_{1} \omega(k_{1})^{-1} \cdot  
\int_{-\infty}^{+\infty} dk_{2} \omega(k_{2})^{-1}  \nonumber \\  \label{w22}  \\ 
&\cdot & \int_{-\infty}^{+\infty} dk_{3} \omega(k_{3})^{-1} \cdot
\int_{-\infty}^{+\infty} dk\, ' | \tilde{g}(k\, ',t)|^{2} \omega \left(
k\, '-{\scriptstyle \sum\limits_{i=1}^3
 } k_i  \right)^{-1} \nonumber
\eea
by the change of variable $k\, '={\scriptstyle \sum\limits_{i=1}^3
 } k_i$. Introducing further the variables $K_{1}, K_{2}, K_{3}$ such that
\begin{eqnarray*}
K_1 &=& k_1+k_2+k_3 \\
K_2 &=& k_1+k_2 \\
K_3 &=& k_1 \; 
\end{eqnarray*}
so that $k_{3}=K_{1}-K_{2}$ and $k_{2}=K_{2}-K_{3}$, we write
(\ref{w22}) as
\be\label{wom}
|| W(\cdot , t) ||^{2}_{2}=\left( \omega^{-1} * \left( \omega^{-1} * \left( 
\omega^{-1} * \left( \omega^{-1} * | \tilde{g} |^{2} \right)
\right) \right) \right) (0),
\ee
where the convolution is defined as usual by
\[
\left (f*g \right )(k)=\int_{-\infty}^{+\infty} dk_{1} f(k-k_{1})g(k_{1}).
\]
Consider, now, the quantity associated to the right-hand side of (\ref{w22}):
\ba
I(q, t)&\equiv& \int_{-\infty}^{+\infty} dk_{1}\; \omega(k_{1}-q)^{-1} 
\cdot  
\int_{-\infty}^{+\infty} dk_{2}\; \omega(k_{2})^{-1} \nonumber \\ \label{I} \\
&\cdot &
\int_{-\infty}^{+\infty} dk_{3} \;\omega(k_{3})^{-1} \cdot
\int_{-\infty}^{+\infty} dk\, ' \; | \tilde{g}(k\, ',t)|^{2} \omega \left(
k\, '- {\scriptstyle \sum\limits_{i=1}^3
 } k_i \right )^{-1} \nonumber
\eea
Since $g\in {\cal D}(\mathbb{R})$ this function is 
differentiable, hence continuous, in $q$ for any compact subset 
containing the origin, which implies that $I(0,t) \leq \|I(\cdot 
,t)\|_{\infty}$ (where $\| \cdot \|_{\infty}$-norm is with respect to the 
$q$-variable).

We now apply Young's inequality \cite{Lieb}
\[
|| f*g ||_{r} \leq  C_{rpq}||f||_{p} ||g ||_{q}
\]
with $C_{rpq}$ a constant and
\[
\frac{1}{p}+\frac{1}{q}=1+\frac{1}{r}
\]
to (\ref{wom}), starting with $r=\infty$. Above,
\[
||f||_{p}=\left ( \int_{-\infty}^{+\infty} dk | f(k) |^{p} \right )^{1/p}.
\]
We thus obtain 
\[
\left\| W(\cdot , t) \right\|^{2}_{2} \leq  C_{2 r_1 r_2}\left\| \omega^{-1} \right\|_{r_{1}}\;
\left\| \left( \omega^{-1} * \left( \omega^{-1} * \left( \omega^{-1} * 
| \tilde{g} |^{2} \right) \right) \right) \right\|_{r_{2}} 
\]
with $r^{-1}_{1}+r^{-1}_{2}=1$, and so on, up to (indicating all the constants resulting from the Young's inequality by $C'$)
\be\label{wr8}
|| W(\cdot , t) ||^{2}_{2} \leq  C' \left\| \omega^{-1} \right\|_{r_{1}}
\; \left\| \omega^{-1} \right\|_{r_{3}}\;  \left\| \omega^{-1} \right\|_{r_{5}}
\; \left\| \omega^{-1} \right\|_{r_{7}} \; \left\| |\tilde{g} |^{2} \right\|_{r_{8}} 
\ee
with $r^{-1}_{3}+r^{-1}_{4}=1+r^{-1}_{2}$, 
$r^{-1}_{5}+r^{-1}_{6}=1+r^{-1}_{4}$,
$r^{-1}_{7}+r^{-1}_{8}=1+r^{-1}_{6}$. We require $r_{i} > 1$, for 
$i=1,3,5,7$, so that $|| \omega^{-1} ||_{r_{i}} < \infty$, the
choice $r_{1}=r_{2}=2$, $r_{3}=r_{4}=\frac{4}{3}$, 
$r_{5}=r_{6}=\frac{8}{7}$, $r_{7}=r_{8}=\frac{16}{15}$ is, for instance, 
possible. By (\ref{wr8})
\be\label{wgr} 
|| W(\cdot , t) ||^{2}_{2} \leq C \left\| |\tilde{g} |^{2} \right\|_{r}
\ee
with
\be\label{r}
r > 1 \; .
\ee
Above
\be\label{gtr}
\left\| \; |\tilde{g} |^{2}\; \right\|_{r}=
\left( \int_{-\infty}^{+\infty}  dk \left| 
\tilde{g} (k,t) \right|^{2r} \right)^{1/r}.
\ee
obtaining finally, (\ref{wgr}). \hfill $\square$

{\bf Proof of \ref{main}} By (\ref{gjdes}),
\be\label{NVN}
\left\| \left(N+{\bf 1} \right)^{-1}V_{g}(t) \left(N+{\bf 1} \right)^{-1} \right\| \leq {\rm 
const.} || W ||_{L^{2}}
\ee
and, by (\ref{omega}), $\omega(k) \geq m {\bf 1}$; hence 
\[
\left\| \left( H_{0}+{\bf 1} \right)^{-1}\left( N+{\bf 1} \right)\right\| 
\leq d_{1} \;\;\;\;\;\;\;\;\;\;\;\; \left\| \left( N+{\bf 1} \right)\left( 
H_{0}+{\bf 1} \right)^{-1}\right\| \leq d_{2},
\]
for constants $d_{1}$ e $d_{2}$. Hence, by (\ref{NVN}) and (\ref{w2}),
\be\label{HVH}
\left\| \left( H_{0}+{\bf 1} \right)^{-1}V_{g}(t) \left( H_{0}+{\bf 1} 
\right)^{-1} \right\| \leq {\rm const.} || g( \cdot, t )||_{r}
\ee
with $r > 1$: a fortiori this holds for $H_{g}(\cdot )$ by (\ref{hg}), hence
\be\label{HHH}
\left\| \left( H_{0}+{\bf 1} \right)^{-1}H_{g}(t) \left( H_{0}+{\bf 1} 
\right)^{-1} \right\| \leq {\rm const.} || g( \cdot, t )||_{r}.
\ee
By (\ref{HBH}) and theorem \ref{texist} we only need to prove that the 
l.h.s.  of (\ref{HHH}) is three times differentiable. We shall prove that 
\be
\label{hgo}
\left\| \left( H_{0}+{\bf 1} \right)^{-1}\!
\left( \! \frac{H_{g}(t+h)-H_{g}(t)}{h} -{H '}_{g}(t)\! \right)\! 
\left( H_{0}+{\bf 1} \right)^{-1} \right\| \longrightarrow 0 \;\;\;\;
{\rm as} \;\;\;\; h \rightarrow 0,
\ee
where 
\be
\label{hg'}
H'_{g}(t)=H_{0}+V_{g'}(t)
\ee
with
\be
\label{vg'}
V_{g'}(t)=\int dx : \phi^{4}(x,0) : g'(x,t)
\ee
and
\[
g'(x,t)\equiv\frac{\partial g(x,t)}{\partial t}.
\]
We now prove (\ref{hgo}). By (\ref{HHH}) 
\begin{eqnarray}
J & \equiv & \left\| \left( H_{0}+{\bf 1} \right)^{-1}\!
\left( \! \frac{H_{g}(t+h)-H_{g}(t)}{h} -{H '}_{g}(t)\! \right)\! 
\left( H_{0}+{\bf 1} \right)^{-1} \right\| \nonumber\\
& \leq & {\rm const.} \left[ \!
\int_{-\infty}^{\infty}\!\!\!\!\! dk \left |
\int \!\! dx \, e^{-ikx}\! \left( \! \frac{g(x,t+h)-g(x,t)}{h}
-g'(x,t)\! \right) \right |^{r}\right]^{\frac{1}{r}}\!\!\!.   \label{h2r}
\end{eqnarray}
We now write the integral on the right-hand side of (\ref{h2r}) as
\[
\int_{-\infty}^{\infty} dk \{\cdots \}  =
\int_{-\infty}^{1} dk \{ \cdots \} +
\int_{-1}^{1} dk \{ \cdots \} + 
\int_{1}^{\infty} dk \{\cdots \}  
\]
and estimate the last integral above
\begin{eqnarray}\label{J}
J_+ &\equiv&\int_{1}^{\infty} dk \left |
\int dx e^{-ikx} \left ( \frac{g(x,t+h)-g(x,t)}{h}
-g'(x,t) \right ) \right |^{r} \nonumber\\
& \leq & \int_{1}^{\infty} \frac{dk}{k^{2r}} 
\left | \int  dx e^{-ikx} \left ( 
\frac{\partial^{2}_{x}g(x,t+h)-\partial^{2}_{x} g(x,t)}{h} 
-\partial^{2}_{x}g'(x,t) \right ) \right |^{r}
\end{eqnarray}
where we have used two partial integrations and $\partial_{x}\equiv \frac{\partial 
}{\partial x}$. Let now
\be\label{V}
V(x,t)\equiv \partial^{2}_{x}g(x,t).
\ee
Now $V$ is also an infinitely differentiable function of compact support and
\be\label{Vtay}
V(x, t+h)=V(x,t)+hV^{'}(x,t)+\frac{h^{2}}{2!}V^{''}(x,t+t^{*}_{h}(x))
\ee
by Taylor's formula with remainder, where $0 < t^{*}_{h}(x) < h$. Putting
(\ref{Vtay}) into (\ref{J}) we get
\[
J_+ \leq c'_r h^r \left (
\int_{-\infty}^{\infty}  dx \left | V^{''}(x, t+t^{*}_{h}(x)) \right | 
\right )^{r} \leq c_r \; h^r ( \sup_{x,t} | V^{''}(x,t) | )^r \; ,
\]
where $c'_r$ and $c_r$ are constants depending on $r$. The estimate of $J_- \equiv \int_{-\infty}^{-1} \{\cdots \}$ is similar. The estimate of $J_1 \equiv \int_{-1}^{1} \{\cdots \}$ follows along the same lines, but in this case we should not introduce the partial integrations in order to avoid divergences at $k=0$. Then, we obtain
$$
J\leq {\rm const.} h \left[ A_r ( \sup_{x,t} | g^{''}(x,t) | )^r +B_r ( \sup_{x,t} | V^{''}(x,t) | )^r\right]^{\frac{1}{r}}.
$$
with $A_r$ and $B_r$ constants depending on $r$. Then we have (\ref{hgo}). 

We now notice that the 
bounds (\ref{HHH}) \underline{continue to hold} for $H'_{g}(t)$ with
$|| g(\cdot , t) ||_{r}$ replaced by 
$|| g'(\cdot , t) ||_{r}$ on the right-hand side of (\ref{HHH}). Thus the 
same proof applies to $H'_{g}(t)$, $H''_{g}(t)$, ... and in fact 
$H_{g}(t)$ is infinitely differentiable as an operator from ${\cal F}_{+2}$
to ${\cal F}_{-2}$.
\hfill
$\square$

\begin {proposition}\label{caus}
The $S(g)$ matrix for the $(:\phi^4)_2$ theory, as defined in (\ref{sg}), is unitary and it satisfies the causality condition for disjoint supports [condition $(ii.a)$ -- section \ref{intro}].
\end{proposition}

{\it Proof.} The unitarity follows directly from the existence theorems. For the proof of causality it is convenient explicitly dispose the dependence of the propagators on the function $g$. Let ${\rm supp}_t \; g_1 > {\rm supp}_t \; g_2$ and suppose ${\rm supp}_t \; g_1 \subset (r, +\infty )$ and ${\rm supp}_t \; g_2 \subset (-\infty , r)$, where ${\rm supp}_t$ stands for the support in the time variable. Then, for $t>r>s$ we have
\be\label{utrs}
U^D_{(g_1+g_2)}(t,s) = U^D_{(g_1+g_2)}(t,r) U^D_{(g_1+g_2)}(r,s)
\ee
but
\begin{eqnarray*}
i\frac{\partial }{\partial t} U^D_{(g_1+g_2)}(t,r)\Psi &=& H^D_{(g_1+g_2)}(t) U^D_{(g_1+g_2)}(t,r)\Psi \\
&=&H^D_{g_1}(t) U^D_{(g_1+g_2)}(t,r)\Psi
\end{eqnarray*}
and, by the uniqueness of the solutions of the above equation, we have 
$ U^D_{(g_1+g_2)}(t,r)= U^D_{g_1}(t,r)$. Analogously, we have 
$ U^D_{(g_1+g_2)}(r,s)= U^D_{g_2}(r,s)$. This, together with (\ref{utrs}) imply that 
$$
U^D_{(g_1+g_2)}(t,s) = U^D_{g_1}(t,r) U^D_{g_2}(r,s)
$$
from this equation and the fact that $ U^D_{g_1}(t,s) = U^D_{g_1}(t,r)$ and $ U^D_{g_2}(r,s) = U^D_{g_2}(t,s)$ due to the support properties of $g_1$ and $g_2$, we finally have
\be\label{utrs1}
U^D_{(g_1+g_2)}(t,s) = U^D_{g_1}(t,s) U^D_{g_2}(t,s)
\ee
Then, by (\ref{utrs1}) and the definition (\ref{sg}), we obtain
$$
S(g_1+g_2) = S(g_1) S(g_2)\; , 
$$ 
\hfill $\square$


\section{The Relation Between Kisy\'nski's Theory and Theorem \ref{main}} \label{kis}
\setcounter{equation}{0}

Let us now briefly summarize (without proof) some steps in 
Kisy\'nski's proof of theorem \ref{texist}. First of all, we will 
state a crucial auxiliary theorem. Let $X$ be a Banach space with the 
norm $\| \cdot \|$ and $A(t)$, $t\in [-T_1, T_2]$ ($T_1, T_2 >0$), a 
family of linear operators in $X$. Consider the following conditions:
\begin{itemize}

\item[(a)] there exists a family $\| \cdot \|_t$, , $t\in [-T_1, T_2]$,  
of norms in $X$ equivalent to $\| \cdot \|$ such that $\left|\, \|\Psi 
\|_t- \| \Psi \|_s \, \right| \leq k \, \| \Psi \|_s \, |t-s|$ with $k= 
{\rm const.}$, $-T_1\leq s,t\leq T_2$ and $\Psi \in X$;
 
\item[(b)] for all $t\in [-T_1, T_2]$ the set $D (A(t))$ is dense in $X$;

\item[(c)] there\hfill exists\hfill a\hfill constant\hfill $\lambda_0\geq 
0$\hfill such\hfill that\hfill $R(\lambda - \epsilon A(t)) = X$\hfill 
and\hfill \\ $\left\| (\lambda -\epsilon A(t))\Psi \right\|_t \geq 
(\lambda -\lambda_0)\| \Psi \|_t$ for $\epsilon =\pm 1$, $\lambda > 
\lambda_0$, $t\in [-T_1, T_2]$ and $\Psi \in D(A(t))$;

\item[(d)] there exists a family $R(t)$, $t\in [-T_1, T_2]$, of 
invertible bounded linear operators in $X$, such that $R(t)$ is twice 
weakly continuously differentiable in $[-T_1, T_2]$ and $\left( 
R(t)\right)^{-1} D (A(t)) = Y = {\rm const.}$ $\forall t\in [-T_1, T_2]$;

\item[(e)] $\left( R(t)\right)^{-1}A(t)R(t)$ is weakly continuously 
differentiable.

\end{itemize}
Above $R(A)$ stands for the range of the operator $A$. Then we have:
\begin{axiom}\label{k4.4}{\rm (\cite{Kis64}, Theorem 4.4)}
Let the conditions (a) - (e) be satisfied. Then there exists a  
two-parameter family of propagators $U(t,s)$, $-T_1\leq s,t \leq T_2$, 
such that 
$$
\Psi (t) \equiv U(t,s) \Psi (s) \; , \hspace{1.0cm} \Psi (s) \in D (A(s)) \; ,
$$

\noindent
is the unique solution of the problem 
\be
\label{eq}
\frac{d}{dt} \Psi (t)= A(t) \Psi (t)
\ee

\noindent
with initial data $\Psi (s)$. The bounded propagators $U(t,s)$ are 
strongly continuous on $-T_1\leq s,t \leq T_2$ and satisfy:
\begin{eqnarray}
&&U(t,t) =1\; , \hspace{3.7cm} \forall \;  t\in [-T_1, T_2]\; ; \label{U1} \\ 
&&U(t,s)U(s,r) =U(t,r)\; , \hspace{1.5cm} {\rm for} -T_1\leq r,s,t\leq T_2\; ; \label{UUU} \\ 
&&U(t,s) D (A(s)) = D (A(t))\; , \hspace{1.0cm} {\rm for} -T_1\leq s,t\leq T_2\; ; \label{UDD}
\end{eqnarray}

\noindent
besides, $\forall \; s \in [-T_1, T_2]$ and $\Psi \in D (A(s))$ the 
function $U(t,s)\Psi$ is continuously differentiable (in the sense of the 
norm) in $X$, satisfying:
\be\label{eqU}
\frac{d}{dt} U(t,s) \Psi = A(t) U(t,s) \Psi \; .
\ee
\end{axiom}

The method of proof of this theorem is to reduce the problem to the case 
where we have an operator with constant domain by making use of the 
properties of $R(t)$ [for an outline of Kisy\'nski's solution of the 
problem (\ref{eq}) with $D(A(t)) = {\rm const.}$ see Appendix A]. 

Let us now consider Kisy\'nski's approach to the abstract 
Schr\"odinger equation
\be\label{sch}
\frac{d}{dt} \Psi (t)= -i A(t) \Psi (t)\; , \hspace{1.0cm} -T_1\leq t \leq T_2
\ee
where $\Psi \in {\cal H}$, with ${\cal H}$ a Hilbert space and $A(t)$ an 
operator in ${\cal H}$ defined as follows. Consider the condition:
\begin{itemize}

\item[($i$)] Let ${\cal H}$ be a Hilbert space, ${\cal H}_+$ a dense 
subset of ${\cal H}$ and, $\forall \, t \in [-T_1,T_2]$, let $\langle 
\cdot , \cdot \rangle_t^+$ be a scalar product defined on ${\cal H}_+$ 
which makes it a Hilbert space ${\cal H}_+^t$ algebraically and 
topologically contained in ${\cal H}$. Assume that $\langle \cdot , \cdot 
\rangle_t^+$ is $n$ times ($n\geq 1$) continuously differentiable on 
$[-T_1,T_2]$.
\end{itemize}
If condition ($i$) is satisfied we have
\begin{lemma}
{\rm (\cite{Kis64}, Lemma 7.2)}
The equality
\be
\langle \Phi , \Psi \rangle_t^+ = \langle \Phi , Q(t)\Psi 
\rangle_{-T_1}^+ \; , \hspace{1.0cm} \Phi, \Psi \in {\cal H}_+ \; , \;\; 
t \in [-T_1,T_2]
\ee
defines a bounded $n$ times weakly continuously differentiable operator 
$Q(t)$ on ${\cal H}_+^{-T_1}$. For all fixed $ t \in [-T_1,T_2]$, $Q(t)$ is hermitian with $\inf Q(t) >0$ in ${\cal H}_+^{-T_1}$.
\end{lemma}
Another consequence of condition ($i$) is that we can define an
operator $J_{-T_1}(t)$ by means of the equality (\cite{Kis64}, 
Lemma 7.4)
\be
\langle \Phi , \Psi \rangle = \langle \Phi , J_{-T_1}(t)\Psi \rangle_t^+\; \hspace{1.0cm} \Phi \in {\cal H}_+\, ,\; \Psi \in {\cal H}
\ee
with $J_{-T_1}(t)$ a positive hermitian operator in ${\cal L}({\cal H})$ such that $J_{-T_1} (t) {\cal H}_+$ is a dense subset of ${\cal H}_+^t$. Then, defining 
\be
\left\| \Psi \right\|_t^- \equiv \left\| J_{-T_1}(t) \Psi \right\|_t^+\; , \hspace{1.0cm} \Psi \in {\cal H}\; ,
\ee
it follows that the completion ${\cal H}^t_-={\cal H}^{-T_1}_-\equiv {\cal 
H}_-$ of ${\cal H}$ in the norm $\| \cdot \|_t^-$ contains ${\cal H}$ 
algebraically and topologically (\cite{Kis64}, Lemma 7.5).

Finally, we can define an operator $A(t)$ by means of the form $\langle  
\cdot, \cdot \rangle_t^+$ according to the following lemma:
\begin{lemma}
{\rm (\cite{Kis64}, Lemma 7.7)}
For all $ t \in [-T_1,T_2]$ 
\be
D(A(t)) = \left\{  \Psi \in {\cal H}_+: \; \sup_{\scriptscriptstyle 
\Phi\in {\cal H}_+\, ,\, \|\Phi\| \leq 1} \{ \left| \langle  \Phi, 
\Psi\rangle_t^+\right| \} <+\infty  \right\} 
\ee
\be
\langle \Phi, A(t)\Psi \rangle \equiv \langle  \Phi, \Psi\rangle_t^+\; , 
\hspace{1.0cm} \Psi \in D(A(t))
\ee
define an inversible self-adjoint positive operator $A(t)$ in ${\cal H}$, with
\be
D(A(t))= \left( Q(t) \right)^{-1} D(A(-T_1))
\ee
and
\be
A(t) = \left( J_{-T_1}(t) \right)^{-1} = A(-T_1)Q(t)\; .
\ee
\end{lemma}
Then the operator $A(t)$ is shown to satisfy the Schr\"odinger equation 
(\ref{sch}) and the propagators of problem (\ref{sch}) satisfy the 
properties enumerated in theorem \ref{texist} (\cite{Kis64}, Theorem
8.1). In order to prove his Theorem 8.1 for the operator $A(t)$, as 
defined above, Kisy\'nski made use of theorem \ref{k4.4} identifying 
$R(t)= \left( Q(t)\right)^{-1}$. Let us now show that the $(:\phi^4:)_2$ 
theory satisfies the necessary conditions for theorem \ref{texist}. 
In fact, all we need to show is that condition ($i$) is satisfied. 
However for the benefit of clarity we will explicitly display the main 
operators introduced in Kisy\'nski's proof and some of its 
properties.

The Hilbert space ${\cal H}$ in $(i)$ should be identified with the symmetric Fock 
space ${\cal F}$ (as defined in section \ref{existence}) and ${\cal F}_{+2}=D(H_0)$ is a dense subset of ${\cal F}$. Then, 
taking the closure ${\cal F}_{+2}^t$ of ${\cal F}_{+2}$ in the norm 
induced by the scalar product $\langle\cdot ,\cdot\rangle_t^+$, which is 
related to the operator $\tilde{H}(t)$ [see equation (\ref{hmc})] by 
means of the form (\ref{forma}), i.e., 
\be\label{pesc}
\langle  \Phi, \Psi\rangle_t^+ \equiv S(\Phi, \Psi) 
= \langle \tilde{H}(t)^{1/2}\Phi, \tilde{H}(t)^{1/2}\Psi \rangle
\ee
we can show the following: 
\begin{proposition} 
${\cal F}_{+2}^t$ is a Hilbert space such that
\be\label{ftf}
{\cal F}^{t}_{+2} \subset {\cal F}
\ee
algebraically and topologically. 
\end{proposition}
\begin{itemize}
\item[] {\bf Proof.} That ${\cal F}_{+2}^t$ is a Hilbert space follows 
immediately from the fact that the form defined in (\ref{pesc}) is closed 
(see, e.g., \cite{Far75}). The property that ${\cal F}^{t}_{+2} \subset 
{\cal F}$ algebraically is trivial. So, it remains to show that 
(\ref{ftf}) holds topologically. This is achieved by showing that for $\{ 
f_n\}_{n=1}^{\infty} \in {\cal F}_{+2}$ and $f \in {\cal 
F}_{+2}$ such that
\be\label{conv}
\| f_{n}-f \| \longrightarrow 0  
\ee
we have
\[
\| f_{n}-f \|_{t}^{+} \longrightarrow 0.
\]
To show this, set
\begin{eqnarray}
\left( \| f_{n}-f \|_{t}^{+}\right)^2 &=&\langle \left (f_{n}-f \right ),
\left (f_{n}-f \right ) \rangle _{t}^{+}  \nonumber\\ 
&=& \langle \left (f_{n}-f \right ), \tilde{H}(t) 
\left (f_{n}-f \right ) \rangle  \nonumber\\ 
&=&\langle \left (H_{0}+{\bf 1} \right )
\left (f_{n}-f \right ), \left (H_{0}+{\bf 1} \right 
)^{-1}\tilde{H}(t) \left( H_{0}+{\bf 1}\right)^{-1} \nonumber \\
&\times& \left( H_{0}+{\bf 1}\right)
\left (f_{n}-f \right ) \rangle  \nonumber
\end{eqnarray}
The Schwarz inequality applied to the last term above yields 
\[
\| f_{n}-f \|_{t}^{+} \leq 
\| \left (H_{0}+{\bf 1} \right )^{-1}
\tilde{H}(t) \left (H_{0}+{\bf 1} \right )^{-1} \| 
\;  \| \left (H_{0}+{\bf 1} \right ) \left (f_{n}-f \right ) \|^2
\]
The first term on the right-hand side is bounded due to (\ref{HHH}). 
The second term on the right-hand side converges since
$H_{0}+{\bf 1}$ is a self-adjoint operator (hence closed) and, by hypothesis, (\ref{conv}) holds. Then the proof of the proposition is complete.
\hfill $\square$
\end{itemize}

In addition, it follows straightforwardly from (\ref{pesc}) and theorem 
\ref{main} that $\langle\cdot ,\cdot\rangle_t^+$ is $n$ times 
(infinitely, in fact) continuously differentiable. Then it is proved that 
condition ($i$) is satisfied and theorem \ref{texist} follows as 
proved in \cite{Kis64} and summarized above.

Now we turn to explicitly show the properties of $Q(t)$ in our case. From (\ref{pesc}) and the definition 
\[
\langle \Phi, \Psi \rangle_{t}^{+} \equiv 
\langle \Phi, Q(t) \Psi \rangle^+_{-T_1}
\]
we obtain that $Q(t)$ is the operator
\be\label{Q}
Q(t)=\left ( \tilde{H}(-T_1) \right )^{-1}\tilde{H}(t)
\ee
\begin{proposition}
$Q(t)$, as defined in (\ref{Q}), is a (strictly) positive hermitian 
operator in ${\cal F}_{+2}$ and it is infinitely weakly differentiable.
\end{proposition}
\begin{itemize}
\item[] {\bf Proof.} It follows directly from the properties of the 
scalar product $\langle\cdot ,\cdot\rangle_t^+$ that $Q(t)$ is infinitely 
weakly differentiable.

For $\Phi$, $\Psi \in {\cal F}_{+2}$,  we have
\begin{eqnarray}
\left( \langle \Phi, Q(t) \Psi  \rangle^{+}_{-T_1} \right)^{*} &=&
\langle Q(t)\Psi, \Phi \rangle^{+}_{-T_1} \nonumber \\
&=& \langle \tilde{H}(-T_1)^{-1}\tilde{H}(t)\Psi, \tilde{H}(-T_1) \Phi 
\rangle  = \langle \tilde{H}(t)\Psi, \Phi \rangle, 
\end{eqnarray}
where we have used (\ref{Q}) . We then have that
\begin{eqnarray}
\left (\langle \Phi, Q(t) \Psi \rangle^{+}_{-T_1} \right )^{*}&=&
\langle \Psi, \tilde{H}(-T_1)\left (\tilde{H}(-T_1) \right 
)^{-1}\tilde{H}(t)\Phi \rangle \nonumber\\
&=& \langle \Psi, \tilde{H}(-T_1) Q(t)\Phi \rangle=\langle \Psi, Q(t) \Phi 
\rangle^{+}_{-T_1},
\end{eqnarray}
which proves that $Q(t)$ is hermitian.

In order to prove that $Q(t)$ is strictly positive on ${\cal F}_{+2}$, we 
must remember that, since ${\cal F}_{+2}^t \subset {\cal F}_{+2}$ 
$\forall \; t$ algebraically and topologically, it follows that the norms 
$\|\cdot \|_{-T_1}^+ $ and $\| \cdot \|_t^+$ are equivalent, i.e., there 
exists $a_t\geq 1$ such that $a_t^{-1}\| \cdot\|_{-T_1}^+ \leq \| \cdot\|_t^+ 
\leq a_t\|\cdot \|_{-T_1}^+$. Then, for $\Psi \in {\cal F}_{+2}$, 
\ba
\langle \Psi , Q(t)\Psi \rangle_{-T_1}^+  &=& ( \|\Psi\|^+_t)^2 \nonumber \\
&\geq & a_t^{-2} ( \|\Psi\|^+_{-T_1})^2
\eea
from which it follows that $\inf Q(t) >0$ and the proof is complete.
\hfill $\square$ 
\end{itemize}

\section{Conclusion: Open Problems}
\label{conclusion}
\setcounter{equation}{0}

The problem of the nonperturbative construction of $S(g)$ for the $\left 
(: \phi^{4} : \right )_{2}$ quantum field theory was addressed in 
\cite{Wre72} using Yosida's approach, which requires that the domain of 
$H_{g}(t)$ be time-independent. For test functions $g(x,t)=h_{1}(x) \cdot 
f_{1}(t)$, i.e., of the product form, this condition is satisfied, but 
already for a sum of two products, e.g., $g(x,t)= 
h_{1}(x) \cdot f_{1}(t)+h_{2}(x) \cdot f_{2}(t)$, with $f_{1}$ and $f_{2}$
having disjoint supports, this is no longer true, and thus the results of 
\cite{Wre72} are incomplete. The present approach does not suffer from this 
inconvenience, and $g$ is allowed to be an arbitrary infinitely 
differentiable function of compact support. Moreover, the use of a scale 
of spaces makes the theory very flexible, being applicable to more 
singular super-renormalizable theories, as well as to four-dimensional 
theories with an ultra-violet cutoff. It is a very challenging problem to 
discover a possibility of ``renormalization" of the exponentials of the 
type (A.7) in the latter, in analogy to the interesting approach of Barata 
\cite{Bar00} and Gentile \cite{Gen02} to the study of certain two-level 
systems.

There are, however, open problems even to finish this program for the 
present $\left (: \phi^{4} : \right )_{2}$ theory: proof of causality for 
space-like supports ({\it ii. b)} and proof of Lorentz covariance ({\it 
iii}). For this purpose, the method outlined in \cite{Wre72} seems natural: 
the above properties would follow from a proof of Faris's product formula 
\cite{Far67} under the assumptions of Theorem IV 1. We shall return to 
this problem in the future.

\section*{\large Acknowledgments}

One of us (W.F.W.) thanks Prof. K. Hepp for posing this problem in 1971 and W.F.W. and L.A.M. thank Prof. K. Fredenhagen for fruitful discussions. W.F.W. was supported in part by CNPq. L.A.M. was supported by FAPESP 
under grant 99/04079-1. O.B. greatly appreciates the financial support 
by Fapesp under grant 01/08485-6.


\renewcommand{\thesection}{\Alph{section}}
\setcounter{section}{1}
\setcounter{axiom}{0}
\section*{Appendix A}
\label{apa}
\newcounter{apend}
\setcounter{apend}{1}
\renewcommand{\theequation}{\Alph{apend}.\arabic{equation}}
\setcounter{equation}{0}

Let us consider the problem (\ref{eq}) for the case in which $D(A(t))= 
{\rm const.}$. The notation is as in the first part of section \ref{kis}.

Consider the following conditions (in what follows $t\in [-T_1,T_2]$, 
unless otherwise specified):
\begin{itemize}
\item[(i)] there exists a family $\|\cdot \|_t$, of norms in $X$ such that $a^{-1}\| \Psi \| \leq \| \Psi\|_t\leq \|\Psi \|_s \leq a \| \Psi \|$, $a\geq 1$, for $-T_1\leq s\leq t\leq T_2$ and $\Psi \in X$;

\item[(ii)] $Y$ is a dense subset of $X$ with $D(A(t))=Y$;

\item[(iii)] for all $\lambda > 0$ and $\Psi \in Y$ we have $R(\lambda - A(t)) =X$ and $\|(\lambda - A(t))\Psi \|_t\geq \lambda \| \Psi \|_t$; 

\item[(iv)] $A(t)$ is weakly continuously differentiable.

\end{itemize}

\begin{axiom}
{\rm (\cite{Kis64}, theorem 3.0)} Let the conditions $({\rm i})$ --$({\rm 
iv})$ be satisfied. Then, there exists a unique solution of the problem 
(\ref{eq}) and the corresponding propagator $U(t,s)$ is strongly 
continuous in $-T_1\leq s\leq t\leq T_2$ and satisfies the properties 
(\ref{U1}), (\ref{UUU}), (\ref{UDD}) and (\ref{eqU}).
\end{axiom}

Now we shall explain some aspects of Kisy\'nski's proof of this 
theorem. Consider the family of equations 
\be
\label{eqan}
\frac{d}{dt}\Phi (t) =A_n(t)\Phi (t)\; , \hspace{1.0cm}\Phi(0)=\Phi_0 \; 
,\hspace{1.0cm} n=1,2,\cdots \; ,
\ee
with 
\be
\label{an}
A_n(t)=nA(t)\left(n-A(t)\right)^{-1}\; .
\ee
The set $Y$ supplied with the norm $|\! |\! |\cdot |\! |\! |_t = 
\|((1-A(t)) \cdot \|$ is a Banach space algebraically and topologically 
contained in $X$. Then, from (i) and (iv), it follows that $A(t)\in {\cal 
L} (Y, X)$ is a weakly continuously differentiable operator, which, by 
the Banach-Steinhaus theorem, implies $\| A(t)\Phi\| \leq C |\! |\! | 
\Phi |\! |\! |_0$ for $\Phi \in Y$ and some constant $C$ (the equivalence 
of the norms $ |\! |\! |\cdot |\! |\! |_t$ was used). So, by using (i) 
and (iii), it follows that 
\be
\left\| \Phi - n(n-A(t))^{-1} \Phi \right\| = \frac{1}{n} \left\| (1-A(t)/n)^{-1} \left( A(t)\Phi \right) \right\| \leq \frac{C a^2}{n} |\! |\! | \Phi |\! |\! |_0 \; ,
\ee
which implies that $n(n-A(t))^{-1}$ converges strongly and uniformly to 
$1$. Therefore, the sequence of bounded operators $A_n(t)$ converges 
strongly to $A(t)$.
The operators $A_n(t)$ are weakly continuosly differentiable, therefore 
they satisfy a Lipschitz condition in the sense of the norm. Hence, it 
follows that $A_n(t)$ is continuous in the sense of the norm and Yosida's  
method \cite{Yos80} guarantees the existence and the uniqueness of the 
evolution operators $U_n(t,s)$ of equation (\ref{eqan}) satisfying
the properties equivalent to (\ref{U1}) -- (\ref{eqU}). Besides, 
$U_n(t,s)$ satisfy \cite{Kis64}
\be
\label{unm}
\| U_n(t,s)\| \leq M \; .
\ee

Before proceeding we will consider the equation (\ref{eqan}) perturbed by 
the bounded (in $X$) weakly continuous operator 
$B(t)=-\frac{dA(t)}{dt}(1-A(t))^{-1}$, that is,
\be\label{eqpert}
\frac{d}{dt}\Phi (t) = \left( A_n(t)+B(t)\right) \Phi (t) \; , 
\hspace{1.0cm}\Phi(0)=\Phi_0 \; 
\ee
The evolution operator of (\ref{eqpert}), denoted $H_n(t,s)$, is given by
$$
H_n(t,s) = (1 - A(t))U_n(t,s) (1-A(s))^{-1}\; .
$$
Then, it follows that $ H_n(t,s) \in {\cal L}(X)$ is weakly continuously 
differentiable in $-T_1\leq s,t\leq T_2$, satisfying 
\be
\label{hn}
\|H_n(t,s)\| \leq D \; .
\ee
Next we subdivide the segment $[-T_1,T_2]$ into $K$ equal intervals. 
Then, the conditions ($T\equiv T_1+T_2$)
\be
U_{nK}(t,s) =\exp \left\{ (t-s) A_n({\scriptstyle - T_1+\frac{i-1}{K}T}) 
\right\} \; ,
\ee
$- T_1+\frac{i-1}{K}T \leq s,t \leq - T_1+\frac{i}{K}T$, 
$i=1, \ldots , K$, and
\be
U_{nK}(t,s)U_{nK}(s,r) =U_{nK}(t,s)\; , \hspace{1.0cm} -T_1\leq r,s,t\leq 
T_2\; ,
\ee
define a unique family of operators $U_{nK}(t,s) \in {\cal L}(X)$ 
continuous in the sense of the norm such that 
\be
\label{bounduk}
\|U_{nK}(t,s)\| \leq a^2 \; .
\ee
The operators $ U_{nK}(t,s)$ satisfy 
$$
\frac{\partial}{\partial s}U_{nK}(t,s) = -U_{nK}(t,s) A_n({\scriptstyle - 
T_1+ \frac{T}{K}\left[ \frac{Ks}{T}\right] } )\; ,
$$
where $[(Ks)/T]$ stands for the integer part of $(Ks)/T$. Besides, for 
fixed $K$, $U_{nK}(t,s)$, $n=1,2,\ldots$, is a sequence uniformly 
strongly convergent in $ - T_1\leq s\leq t\leq T_2$.

Then, by integrating $\frac{\partial}{\partial \tau}U_{nK}(t,\tau 
)U_n(\tau , s)$ we obtain 
\be
\label{unuk}
U_n(t,s)-U_{nK}(t,s) = \int_s^t U_{nK}(t,\tau ) \left( A_n(\tau )- 
A_n({\scriptstyle - T_1+ \frac{T}{K}\left[ \frac{K\tau }{T}\right] } )
\right) U_n(\tau ,s) d\tau \; .
\ee
We have \cite{Kis64}
\be
\label{aa}
\left\| A_n(\tau )\Phi - A_n({\scriptstyle - T_1+ \frac{T}{K}\left[ 
\frac{K\tau }{T}\right] } ) \Phi \right\| \leq \frac{{\rm const.}}{K} |\! 
|\! | \Phi |\! |\! |_0 \; .
\ee
Then, since
$$
U_n(t,s) = (1-A(t))^{-1} H_n(t,s)(1-A(s))
$$
and $(1-A(s))\in {\cal L}(Y,X)$ and $(1-A(t))^{-1}\in {\cal L}(X,Y)$ are 
weakly differentiable, we obtain, by using (\ref{hn}), 
\be
\label{uf}
|\! |\! | U_n(t,s) \Phi |\! |\! |_0 \leq {\rm const.} |\! |\! | \Phi |\! 
|\! |_0 \; ,
\ee
for $\Phi \in Y$ and $-T_1\leq s\leq t\leq T_2$. Then, from 
(\ref{bounduk}), (\ref{unuk}), (\ref{aa}) and (\ref{uf}), it follows that
\be
\label{uconvu}
\left\| U_n(t,s)\Phi - U_{nK}(t,s)\Phi \right\| \leq \frac{\rm L}{K} |\! 
|\! | \Phi |\! |\! |_0\; ,
\ee
with $L={\rm constant}$.

Now, for $\Phi \in Y$ we have
\begin{eqnarray}
\left\|U_n(t,s)\Phi - U_m(t,s)\Phi \right\| &\leq & \left\|U_n(t,s)\Phi - 
U_{nK}(t,s)\Phi \right\| \nonumber \\
&+& \left\|U_{mK}(t,s)\Phi - U_m(t,s)\Phi \right\| \nonumber \\
&+&\left\|U_{nK}(t,s)\Phi - U_{mK}(t,s)\Phi \right\| \nonumber \\
&\leq & 2\frac{L}{K} |\! |\! |\Phi |\! |\! |_0 +\left\|U_{nK}(t,s)\Phi - 
U_{mK}(t,s)\Phi \right\| \label{UUUU}
\end{eqnarray}
The first term in the r.h.s. may be made arbitrarily small for large $K$. 
After this, one chooses $n$ and $m$ so large that the second term becomes 
arbitrarily small for all $-T_1\leq s\leq t \leq T_2$, since the sequence 
$U_{nK}(t,s)$ is uniformly strongly convergent. Since $Y$ is dense in 
$X$, and from (\ref{unm}), (\ref{UUUU}) implies that the convergence is 
in all of $X$, in the triangle
$-T_1\leq s\leq t \leq T_2$. Then, it follows directly from the 
properties of $U_n(t,s)$ that $U(t,s) = s - \lim_{n\rightarrow \infty} 
U_n(t,s)$ is the evolution operator of (\ref{eq}) for constant domain 
\cite{Kis64}.

{\bf Remark.} The proof outlined above is valid for $ - T_1 \leq s\leq t \leq T_2$. However, by substituting the conditions (i) and (iii) above by the conditions (a) and (c) in the theorem \ref{k4.4} the proof can be extended for the square $-T_1\leq s,t \leq T_2$.


\begin{thebibliography}{99}
\section*{References}


\bibitem{BLT75} N.N. Bogoliubov, A. N. Logunov and I. T. Todorov, 
{\it Introduction to Axiomatic Quantum Field Theory} (Benjamin, 1975).

\bibitem{EGl73} H. Epstein and V. Glaser, {\it Ann. Inst. H. Poincar\'{e} 
A} {\bf 19}, 211 (1973).

\bibitem{Sch01} G. Scharf, {\it Gauge Theories: A True Ghost Story} 
(Wiley, 2001).

\bibitem{DFr99} M. D\"utsch and K. Fredenhagen, {\it Commun. Math. Phys.} 
{\bf 203}, 71 (1999).

\bibitem{OSe76} K. Osterwalder and R. S\'eneor, {\it Helv. Phys. Acta} 
{\bf 49}, 525 (1976).

\bibitem{EEF76} J.-P. Eckmann, H. Epstein and J. Fr\"ohlich, 
{\it Ann. Inst. Henri Poincar\'e} {\bf 25}, 1 (1976).

\bibitem{GJS74} J. Glimm, A. Jaffe and T. Spencer, {\it Ann. Math.} {\bf 100}, 
585 (1974).

\bibitem{EGl76} H. Epstein and V. Glaser, {\it Adiabatic Limit in 
Perturbation Theory}, in ``Renormalization Theory'', G. Velo 
and A. S. Wightman (eds.) (1976).

\bibitem{GJa70} J. Glimm and A. Jaffe, {\it Quantum Field Theory Models}, 
in Les Houches 1970: ``Statistical Mechanics and Quantum Field Theory'', 
C. DeWitt and R. Stora (eds.) (Gordon and Breach, New York, 1972).

\bibitem{GJa72} J. Glimm and A. Jaffe, {\it Boson Quantum Field Models}, 
in ``London 1971, Mathematics of Contemporary Physics'', 
R. F. Streater (ed.) (Academic Press, London, 1972).

\bibitem{Kis64} J. Kisy\'nski, {\it Studia Math.} {\bf 23}, 285 (1964). 

\bibitem{Sim71} B. Simon, {\it Quantum Mechanics of Hamiltonians 
Defined as Quadratic Forms}, (Princeton University Press, 1971).

\bibitem{Yos80} K. Yosida, {\it Functional Analysis}, 6th ed. (Springer-Verlag, 
New York, 1980).

\bibitem{GJa68} J. Glimm and A. Jaffe, {\it Phys. Rev.} {\bf 176}, 
1945 (1968).

\bibitem{Far75} W. G. Faris, {\it Self-Adjoint Operators}, Lecture Notes 
in Math., v. 433 (Springer-Verlag, Berlin, 1975).

\bibitem{Lieb} E. H. Lieb and M. Loss, {\it Analysis} (Am. Math. Soc., 1997).

\bibitem{Wre72} W. Wreszinski, {\it Theor. Math. Phys.} {\bf 11}, 547 (1972). 

\bibitem{Bar00} J. C. A. Barata, {\it Rev. Math. Phys.} {\bf 12}, 25 (2000).

\bibitem{Gen02} G. Gentile, {\it Commun. Math. Phys.} {\bf 242}, 221 (2003).

\bibitem{Far67} W. G. Faris, {\it J. Funct. Anal.} {\bf 1}, 93 (1967).


\end{thebibliography}
\end{document}